\documentclass[11pt]{article}
\usepackage{amssymb}
\def\one{1\hskip -.37em 1}     
\begin{document}
\begin{titlepage}
\begin{centering}
 
{\ }\vspace{2cm}
 
{\Large\bf A General Four-Fermion Effective Lagrangian for}

\vspace{5pt}

{\Large\bf Dirac and Majorana Neutrino-Charged Matter Interactions}

\vspace{2cm}

Jean El Bachir Mendy\\
\vspace{0.5cm}

{\em Department of Physics, University Cheikh Anta Diop}\\
{\em Dakar, Senegal}\\
{\tt mendiz19@yahoo.fr}

\vspace{0.5cm}

and

\vspace{0.5cm}

Jan Govaerts\\
\vspace{1.0cm}
{\em Institute of Nuclear Physics, Catholic University of Louvain}\\
{\em 2, Chemin du Cyclotron, B-1348 Louvain-la-Neuve, Belgium}\\
{\tt jan.govaerts@fynu.ucl.ac.be}

\vspace{1.5cm}

\begin{abstract}

\noindent
Given the most general Lorentz invariant four-fermion effective
interaction possible for two neutrinos and two charged fermions,
whether quarks or leptons, all possible $2\rightarrow 2$ processes 
involving two neutrinos, whether Dirac or Majorana ones, and two charged 
fermions are considered. Explicit and convenient expressions are
given for the associated differential cross-sections. Such a parametrization
should help assess the sensitivity to physics beyond the Standard Model 
of neutrino beam experiments which are in the design stage at neutrino
factories.

\end{abstract}

\vspace{35pt}
 
To be published in the Proceedings of the\\
Second International Workshop on Contemporary Problems in Mathematical
Physics,\\
Institut de Math\'ematiques et de Sciences Physiques (IMSP), Universit\'e
d'Abomey-Calavi,\\
Cotonou, Republic of Benin\\
October 28$^{\rm st}$ - November 2$^{\rm th}$, 2001

\end{centering} 

\vspace{100pt}

%\noindent hep-ph/yymmddd\\
%\noindent July 2002

\end{titlepage}

\section{\bf Introduction}
\label{Sect1}

Neutrino physics is called to play an important role in fundamental physics,
certainly for the decades to come, ranging from high
energy particle physics to astroparticle physics and cosmology. Neutrinos
hold the key to some of the present-day mysteries in these fields, 
with the potential for profound breakthroughs in our search of
the unification of all interactions. However in this endeavour, the lead
is certainly to come from experiments: they must identify which, if
any, of the many theoretical avenues that have been imagined, has actually
been chosen by Nature. Thus ambitious projects
are in the ma\-king, ranging from neutrino telescopes to accelerator
experiments and neutrino factories. 

Given this context, it should be of interest to have available
a general model independent parametrization of large classes of
processes within an effective description relevant for a given energy range. 
For example in the 1950's, such a general four-fermion effective 
interaction\cite{JTW} has enabled the identification of the $(V-A)$ 
structure of the weak interaction in $\beta$-decay and is still used in 
modern precision measurements.\cite{Deutsch} A similar parametrization is 
also of relevance to precision studies in the muon sector looking for 
physics beyond the Standard Model within purely leptonic 
processes.\cite{Fetscher,PDG} Likewise, such a parametrization of
semi-leptonic muonic interactions is available in the intermediate
energy range of nuclear muon capture.\cite{JG} Given the eventual advent 
of neutrino factories, an analogous general parametrization should thus be of
interest, in order to assess the potential of any given experiment based
on neutrino beams to constrain the parameter space of possible physics
beyond the Standard Model (SM) within the neutrino sector.

In this note, we wish to report briefly on such an analysis and some of its
results,\cite{Mendy1,Mendy2} based on the most general four-fermion effective 
interaction possible of two neutrinos and two charged fermions (whether 
leptons or quarks) of fixed ``flavours", or rather more correctly, of 
definite mass eigenstates, solely constrained by the requirements of Lorentz 
invariance and electric charge conservation. For instance, even though
this might be realized only in peculiar classes of models beyond 
the SM, allowance is made for the possibility that both the neutrino fields 
and their charge conjugates couple in the effective Lagrangian density. 
Furthermore, the analysis is developed separately whether for Dirac or 
Majorana neutrinos, with the hope to identify circumstances under which 
scattering experiments involving neutrinos could help discriminate between 
these two cases through different angular correlations for differential 
cross sections, given the high rates to be expected at neutrino factories. 
As is well known, the ``practical Dirac-Majorana confusion theorem" 
states\cite{Kayser1} that within the SM, namely in the limit of massless 
neutrinos as well as $(V-A)$ interactions only, these two possibilities are 
physically totally equivalent, and hence cannot be distinguished. On the 
other hand, relaxing the purely $(V-A)$ structure of the electroweak 
interaction by including at least another interaction whose chirality 
structure is different, should suffice to evade this conclusion, even in the 
limit of massless neutrinos. 

The general classes of processes comprise
neutrino pair annihilation into charged leptons,\footnote{Henceforth, the
charged fermions are referred to as leptons, even though exactly the same
analysis and results apply to quarks, with due account then for the
quark colour degree of freedom and the quark structure of the hadrons involved. 
Also, charged leptons will simply be called leptons, for short.} the inverse 
process of neutrino pair production through lepton annihilation, and finally 
neutrino-lepton scattering. These processes will also be considered whether 
either one or both pairs of neutrino and lepton flavours, $(a,b)$ and $(i,j)$ 
respectively, are identical or not. The sole implicit assumption is that 
the energy available to the reaction is both sufficiently large in order 
to justify ignoring neutrino and lepton masses,
and sufficiently small in order to justify the four-fermion parametrization
of the boson exchanges responsible for the interactions. 
Hence, calculations are performed in the limit of zero mass
for all external neutrino and lepton mass eigenstates.
Nonetheless, effects that distinguish Majorana from Dirac neutrinos 
survive in this limit. Note that this massless approximation
also justifies our abuse of language in referring to neutrino mass eigenstates
as flavour eigenstates.

Sect.\ref{Sect2} provides a list of useful relations for Dirac and
Majorana spinors. Sect.\ref{Sect3} discusses the general four-fermion 
effective Lagrangian used in our ana\-ly\-sis. Sects.\ref{Sect4} to
\ref{Sect6} then list the results for the three classes of processes
mentioned above. Some comments and concluding remarks are made in 
Sect.\ref{Sect7}.

\section{A Compendium of Properties}
\label{Sect2}

\subsection{Dirac, Weyl and Majorana spinors}
\label{Subsect2.1}

This section present facts relevant to Dirac, Weyl and Majorana spinors.
Since all processes are considered in the massless limit,
the representation of the Clifford-Dirac algebra 
$\{\gamma^\mu,\gamma^\nu\}=2g^{\mu\nu}=2\,{\rm diag}\,(+---)$ used throughout 
is the chiral one ($\mu=0,1,2,3$; $i=1,2,3$),
\begin{equation}
\gamma^0=\left(\begin{array}{c c}
0 & -\one \\
-\one & 0 \end{array}\right)\ ,\ 
\gamma^i=\left(\begin{array}{c c}
0 & \sigma^i \\
-\sigma^i & 0 \end{array}\right)\ ,\ 
\gamma_5=i\gamma^0\gamma^1\gamma^2\gamma^3=\left(\begin{array}{c c}
\one & 0 \\
0 & -\one \end{array}\right)\ ,
\end{equation}
$\sigma^i$ being of course the Pauli matrices.
The chiral projectors $P_\eta$ ($\eta=\pm$) are given by
\begin{equation}
P_\eta=\frac{1}{2}\left[1+\eta\gamma_5\right]\ \ ,\ \ 
P^2_\eta=P_\eta\ \ \ ,\ \ \ P_\eta P_{-\eta}=0\ \ ,\ \ \eta=+,-\ .
\end{equation}

By definition, the charge conjugation matrix $C$ is such that
\begin{equation}
\begin{array}{r c l c r c l}
C^{-1}\one C&=&\one^{\rm T}\ &,&\
C^{-1}\gamma_5C&=&\gamma_5^{\rm T}\ ,\\
C^{-1}\gamma^\mu C&=&-{\gamma^\mu}^{\rm T}\ &,&\
C^{-1}\left(\gamma^\mu\gamma_5\right)C&=&
{\left(\gamma^\mu\gamma_5\right)}^{\rm T}\ ,\\
C^{-1}\sigma_{\mu\nu}C&=&-\sigma_{\mu\nu}^{\rm T}\ &,&\
C^{-1}\left(\sigma_{\mu\nu}\gamma_5\right)C&=&
-\left(\sigma_{\mu\nu}\gamma_5\right)^{\rm T}\ ,
\end{array}
\end{equation}
with $C^{\rm T}=C^\dagger=-C$ and $CC^\dagger=\one=C^\dagger C$,
and is realized in the chiral representation by
$C=\,{\rm diag}(-i\sigma^2,\ i\sigma^2)$.

Given a four component Dirac spinor $\psi$, our definition of the associated
charge conjugate spinor is such that
\begin{equation}
\psi_c=\psi^c=\lambda\,C\overline{\psi}^{\rm T}\ ,
\end{equation}
where $\lambda$ is some arbitrary unit phase factor, whose value
may depend on the spinor field. 

Solutions to the free massless Dirac equation may be expanded as follows
in the helicity basis in the case of a Dirac spinor $\psi_D(x)$,
\begin{equation}
\psi_D(x)=\int_{(\infty)}\frac{d^3\vec{k}}{(2\pi)^32|\vec{k}|}\,
\sum_{\eta=\pm}\left[e^{-ik\cdot x}u(\vec{k},\eta)b(\vec{k},\eta)+
e^{ik\cdot x}v(\vec{k},\eta)d^\dagger(\vec{k},\eta)\right]\ .
\end{equation}
Here, the fermionic creation and annihilation operators have the Lorentz
covariant normalization
\begin{equation}
\left\{b(\vec{k},\eta),b^\dagger(\vec{k}',\eta')\right\}=
(2\pi)^32|\vec{k}|\delta_{\eta,\eta'}\delta^{(3)}(\vec{k}-\vec{k}')=
\left\{d(\vec{k},\eta),d^\dagger(\vec{k}',\eta')\right\}\ ,
\end{equation}
while the plane wave spinors $u(\vec{k},\eta)$ and $v(\vec{k},\eta)$ are
given by,
\begin{equation}
u(\vec{k},+)=v(\vec{k},-)=\sqrt{2|\vec{k}|}\left(\begin{array}{c}
\chi_+(\hat{k}) \\ 0 \end{array}\right)\,,\,
u(\vec{k},-)=v(\vec{k},+)=\sqrt{2|\vec{k}|}\left(\begin{array}{c}
0 \\ \chi_-(\hat{k}) \end{array}\right)\,,
\end{equation}
with the Pauli bi-spinors
\begin{equation}
\chi_+(\hat{k})=\left(\begin{array}{c}
e^{-i\varphi/2}\cos\theta/2 \\ e^{i\varphi/2}\sin\theta/2 \end{array}\right)
\ \ ,\ \ 
\chi_-(\hat{k})=\left(\begin{array}{c}
-e^{-i\varphi/2}\sin\theta/2 \\ e^{i\varphi/2}\cos\theta/2 \end{array}\right)\ ,
\end{equation}
such that
$\hat{k}\cdot\vec{\sigma}\,\chi_\eta(\hat{k})=\eta\,\chi_\eta(\hat{k})$ and
$\chi_\eta(\hat{k})\chi^\dagger_\eta(\hat{k})=
(\one+\eta\hat{k}\cdot\vec{\sigma})/2$,
$\varphi$ and $\theta$ being the spherical angles for the
unit vector $\hat{k}=\vec{k}/|\vec{k}|$ with respect to the axes $i=1,2,3$,
namely $\hat{k}=(\sin\theta\cos\varphi,\sin\theta\sin\varphi,\cos\theta)$. 

The value of $\eta=\pm$ coincides with the helicity
of the associated massless one-particle states, as well as
the chirality of the associated quantum field.
Left- or right-handed four component Weyl spinors, with
$\eta=-$ and $\eta=+$ respectively, read
\begin{equation}
\psi_\eta(x)=\int_{(\infty)}\frac{d^3\vec{k}}{(2\pi)^32|\vec{k}|}\,
\left[e^{-ik\cdot x}u(\vec{k},\eta)b(\vec{k},\eta)+
e^{ik\cdot x}v(\vec{k},-\eta)d^\dagger(\vec{k},-\eta)\right]\ ,
\end{equation}
as implied by the identification
\begin{equation}
\psi_\eta(x)=P_\eta\,\psi_D(x)\ .
\end{equation}
Hence, $b^\dagger(\vec{k},\eta)$ and $d^\dagger(\vec{k},\eta)$ are the
creation operators of a particle and of an antiparticule, respectively, each
of helicity $\eta$ and momentum $\vec{k}$.

This identification may also be established from the chiral properties of
the plane wave spinors,
\begin{equation}
\begin{array}{r c l c r c l}
P_\eta\,u(\vec{k},\eta)&=&u(\vec{k},\eta)\ &,&\ 
P_\eta\,u(\vec{k},-\eta)&=&0\ ,\\
P_\eta\,v(\vec{k},\eta)&=&0\ &,&\
P_\eta\,v(\vec{k},-\eta)&=&v(\vec{k},-\eta)\ , \\
\overline{u}(\vec{k},\eta)\,P_\eta&=&0\ &,&\ 
\overline{u}(\vec{k},-\eta)\,P_\eta&=&\overline{u}(\vec{k},-\eta)\ ,\\
\overline{v}(\vec{k},\eta)\,P_\eta&=&\overline{v}(\vec{k},\eta)\ &,&\ 
\overline{v}(\vec{k},-\eta)\,P_\eta&=&0\ , 
\end{array}
\end{equation}
as well as
\begin{equation}
u(\vec{k},\eta)\overline{u}(\vec{k},\eta)=\frac{\one+\eta\gamma_5}{2}/\!\!\!k
\ \ \ ,\ \ \ 
v(\vec{k},\eta)\overline{v}(\vec{k},\eta)=\frac{\one-\eta\gamma_5}{2}/\!\!\!k
\ \ .
\end{equation}
Their properties under charge conjugation are such that
$C\overline{u}^{\rm T}(\vec{k},\eta)=v(\vec{k},\eta)$,
$C\overline{v}^{\rm T}(\vec{k},\eta)=u(\vec{k},\eta)$,
$\overline{v}(\vec{k},\eta)=u^{\rm T}(\vec{k},\eta)C$ and
$\overline{u}(\vec{k},\eta)=v^{\rm T}(\vec{k},\eta)C$,
these relations being specific to the helicity basis.
Charge conjugates of spinors are then given by, say for
a Dirac spinor $\psi_D(x)$,
\begin{equation}
\psi^c_D(x)=\int_{(\infty)}\frac{d^3\vec{k}}{(2\pi)^32|\vec{k}|}\,
\sum_{\eta=\pm}\left[e^{-ik\cdot x}\lambda u(\vec{k},\eta)d(\vec{k},\eta)+
e^{ik\cdot x}\lambda v(\vec{k},\eta)b^\dagger(\vec{k},\eta)\right]\ .
\end{equation}

As opposed to a Dirac spinor
comprised of two independent Weyl spinors of opposite chiralities,
namely one of each of the two fundamental representations of the 
(covering group of the) Lorentz group, $\psi_D(x)=\psi_+(x)+\psi_-(x)$,
a Majorana spinor $\psi_M(x)$ is a four component spinor, thus also covariant 
under Lorentz transformations, but constructed from a single 
Weyl spinor, say of left-handed chirality\footnote{Since charge conjugation 
exchanges left- and right-handed chiralities, the chirality of the basic 
Weyl spinor used in this construction is irrelevant to the definition of a 
Majorana spinor.} $\eta=-$, and which is invariant under charge 
conjugation\footnote{A similar definition starting from a Dirac 
rather than a Weyl spinor might be contemplated, leading then to 
two independent Majorana spinors, each of which is obtained in the manner 
just described from a single distinct Weyl spinor, namely 
$\psi^{(1)}_M=(\psi_D+\psi^c_D)/\sqrt{2}$ and
$\psi^{(2)}_M=-i(\psi_D-\psi^c_D)/\sqrt{2}$, in complete analogy with the
real and imaginary parts of a single complex scalar field as well as the
physical interpretation of the associated quanta as being particles which
are or not their own antiparticles. Specifically, we have
$\psi^{(1)}_M=\psi^{(1)}_-+{\psi^{(1)}_-}^c$ and
$\psi^{(2)}_M=\psi^{(2)}_-+{\psi^{(2)}_-}^c$ with
$\psi^{(1)}_-=(\psi_-+\psi^c_+)/\sqrt{2}$,
$\psi^{(2)}_-=-i(\psi_--\psi^c_+)/\sqrt{2}$, where
$\psi_D=\psi_-+\psi_+$. Setting either $\psi_-$ or $\psi_+$ to zero,
the Weyl spinors $\psi^{(1)}_-$, $\psi^{(2)}_-$ hence also the Majorana
ones $\psi^{(1)}_M$, $\psi^{(2)}_M$ are then no longer independent,
leading back to the construction described in the body of the text.}
\begin{equation}
\psi_M(x)=\psi_-(x)+\psi^c_-(x)\ \ \ ,\ \ \ 
\psi^c_M(x)=\lambda_MC\overline{\psi}^{\rm T}=\psi_M(x)\ ,
\end{equation}
where the possible spinor dependency of the phase factor $\lambda_M$
is now emphasized. Consequently, the mode expansion of a Majorana spinor 
in the helicity basis is,
\begin{equation}
\psi_M(x)=\int_{(\infty)}\frac{d^3\vec{k}}{(2\pi)^32|\vec{k}|}\,
\sum_{\eta=\pm}\left[e^{-ik\cdot x}u(\vec{k},\eta)a(\vec{k},\eta)+
e^{ik\cdot x}\lambda_M v(\vec{k},\eta)a^\dagger(\vec{k},\eta)\right]\ ,
\end{equation}
where the annihilation and creation operators $a(\vec{k},\eta)$
and $a^\dagger(\vec{k},\eta)$ obey the fermionic algebra
\begin{equation}
\left\{a(\vec{k},\eta),a^\dagger(\vec{k}',\eta')\right\}=
(2\pi)^32|\vec{k}|\delta_{\eta,\eta'}\delta^{(3)}(\vec{k}-\vec{k}')\ .
\end{equation}
In terms of the quanta of the basic Weyl spinor used in the construction,
we have the following correspondence (the complex conjugate
of a complex number $z$ is denoted $z^*$ throughout),
\begin{equation}
\begin{array}{r c l c r c l}
a(\vec{k},-)&:&b(\vec{k},-)\ \ \ &;&\ \ \ 
a^\dagger(\vec{k},-)&:&b^\dagger(\vec{k},-)\ \ ,\\
a(\vec{k},+)&:&\lambda_M d(\vec{k},+)\ \ \ &;&\ \ \ a^\dagger(\vec{k},+)&:&
\lambda^*_Md^\dagger(\vec{k},+)\ \ ,
\end{array}
\label{eq:correscreation}
\end{equation}
showing that $a^\dagger(\vec{k},\eta)$ is the creation
operator of a particle of momentum $\vec{k}$ and helicity $\eta$ which 
is also its own antiparticle. The charge conjugation phase factor
$\lambda_M$ is seen to corresponds to the so-callsed 
``creation phase factor".\cite{Kayser2}

\subsection{Differential cross sections}
\label{Subsect2.2}

All $2\rightarrow 2$ processes to be discussed are considered in their 
center-of-mass (CM) frame, with a kinematics of the form
\begin{equation}
p_1+p_2\rightarrow q_1+q_2\ ,
\end{equation}
the quantities $p_{1,2}$, $q_{1,2}$ standing for the four-momenta
of the respective in-coming and out-going massless particles. Given rotational
invariance, and the fact that all particles are of spin 1/2 and of zero mass,
hence of definite helicity, the sole angle of relevance is the CM scattering 
angle $\theta$ between, say, the momenta $\vec{p}_1$ and $\vec{q}_1$. For all 
the reactions listed hereafter, the same order is used for the pairs $(p_1,p_2)$
and $(q_1,q_2)$ of the initial and final particles involved, hence leading 
always to the same interpretation for this angle $\theta$ as being the 
scattering angle between the first particle in each of these two pairs of 
in-coming and out-going states.

For external particles of definite helicity,
the differential CM cross section of all such processes is given by
\begin{equation}
\frac{d\sigma}{d\Omega_{\hat{q}_1}}=\frac{1}{S_f}\frac{1}{64\pi^2\,s}\,
|{\cal M}|^2\ \ \ ,\ \ \ 
\frac{d\sigma}{d\cos\theta}=\frac{1}{S_f}\frac{1}{32\pi\,s}\,
|{\cal M}|^2\ .
\end{equation}
Here, $\sqrt{s}$ is the reaction total invariant energy, with
$s=(p_1+p_2)^2=(q_1+q_2)^2$,
$d\Omega_{\hat{q}_1}$ is the solid angle associated to the outgoing
particle of normalized momentum $\hat{q}_1=\vec{q}_1/|\vec{q}_1|$,
$S_f=2$ or $S_f=1$ depending on whether the two particles---including their
helicity---in the final state are identical or not, respectively, 
and ${\cal M}$ is Feynman's scattering matrix element. 
Our results are listed in terms of the relevant amplitudes ${\cal M}$.

\section{The Four-Fermion Effective Lagrangian}
\label{Sect3}

Given our assumption concerning the energy and mass scales involved,
an effective four-fermion parametrization is warranted, constrained
by the sole requirements of Lorentz invariance and electric charge 
conservation. Since fermion number is not necessarily conserved,
one may equally well couple the neutrino fields and their charge conjugates 
to the charged fermionic fields. For the latter, Dirac fields represent the 
ordinary charged leptons (or quarks) rather than their antiparticules.
It is relative to this choice that the neutrino fields
and their charge conjugates are thus specified.

We shall consider all processes involving
neutrinos or their antineutrinos of definite flavours $a$ and $b$, as well as
leptons or their antileptons of flavours $i$ and $j$, all denoted
as $\nu_a$, $\nu_b$, $\ell^-_i$ and $\ell^-_j$, respectively. 
Hence, the total four-fermion effective Lagrangian in the case of
Dirac neutrinos is (through a Fierz transformation, its expression
may be brought to the charge-retention form),
\begin{equation}
{\cal L}_{\rm eff}=4\frac{g^2}{8M^2}\left[{\cal L}_D+{\cal L}^\dagger_D\right]
\ \ ,\ \
{\cal L}_D={\cal L}_1+{\cal L}_2+{\cal L}_3+{\cal L}_4\ ,
\end{equation}
each separate contribution being given by
\begin{equation}
\begin{array}{r l}
{\cal L}_1=S_1^{\eta_a,\eta_b}\overline{\nu_a}P_{-\eta_a}\ell_i\
\overline{\ell_j}P_{\eta_b}\nu_b\ &+\
V_1^{\eta_a,\eta_b}\overline{\nu_a}\gamma^\mu P_{\eta_a}\ell_i\
\overline{\ell_j}\gamma_\mu P_{\eta_b}\nu_b\ \\
&+\ \frac{1}{2}T_1^{\eta_a,\eta_b}\overline{\nu_a}\sigma^{\mu\nu}
P_{-\eta_a}\ell_i\ \overline{\ell_j}\sigma_{\mu\nu}P_{\eta_b}\nu_b\ ,
\end{array}
\label{eq:1}
\end{equation}
\begin{equation}
\begin{array}{r l}
{\cal L}_2=S_2^{\eta_a,\eta_b}\overline{\nu^c_a}P_{\eta_a}\ell_i\
\overline{\ell_j}P_{\eta_b}\nu_b\ &+\
V_2^{\eta_a,\eta_b}\overline{\nu^c_a}\gamma^\mu P_{-\eta_a}\ell_i\
\overline{\ell_j}\gamma_\mu P_{\eta_b}\nu_b\ \\
&+\ \frac{1}{2}T_2^{\eta_a,\eta_b}\overline{\nu^c_a}\sigma^{\mu\nu}
P_{\eta_a}\ell_i\ \overline{\ell_j}\sigma_{\mu\nu}P_{\eta_b}\nu_b\ ,
\end{array}
\end{equation}
\begin{equation}
\begin{array}{r l}
{\cal L}_3=S_3^{\eta_a,\eta_b}\overline{\nu_a}P_{-\eta_a}\ell_i\
\overline{\ell_j}P_{-\eta_b}\nu^c_b\ &+\
V_3^{\eta_a,\eta_b}\overline{\nu_a}\gamma^\mu P_{\eta_a}\ell_i\
\overline{\ell_j}\gamma_\mu P_{-\eta_b}\nu^c_b\ \\
&+\
\frac{1}{2}T_3^{\eta_a,\eta_b}\overline{\nu_a}\sigma^{\mu\nu}P_{-\eta_a}\ell_i\
\overline{\ell_j}\sigma_{\mu\nu}P_{-\eta_b}\nu^c_b\ ,
\end{array}
\end{equation}
\begin{equation}
\begin{array}{r l}
{\cal L}_4=S_4^{\eta_a,\eta_b}\overline{\nu^c_a}P_{\eta_a}\ell_i\
\overline{\ell_j}P_{-\eta_b}\nu^c_b\ &+\
V_4^{\eta_a,\eta_b}\overline{\nu^c_a}\gamma^\mu P_{-\eta_a}\ell_i\
\overline{\ell_j}\gamma_\mu P_{-\eta_b}\nu^c_b\ \\
&+\
\frac{1}{2}T_4^{\eta_a,\eta_b}\overline{\nu^c_a}\sigma^{\mu\nu}P_{\eta_a}\ell_i\
\overline{\ell_j}\sigma_{\mu\nu}P_{-\eta_b}\nu^c_b\ ,
\end{array}
\end{equation}
an implicit summation over the chiralities $\eta_a$ and $\eta_b$ being
understood of course. It is important to keep in mind
that no summation over the flavour indices $a$ and $b$, nor $i$ and $j$
is implied; all four of these values are fixed from the outset, keeping
open still the possibility that $a$ and $b$ might be equal or not,
and likewise for $i$ and $j$.

The overall normalization factor $4g^2/8M^2$ involves a dimensionless
coupling constant $g$ as well as a mass scale $M$, while the factor $4$
cancels the two factors $1/2$ present in the chiral projection operators 
$P_{\pm\eta_a}$ and $P_{\pm\eta_b}$.
The rationale for this choice of normalization is that in the limit of 
the SM, $g$ is then the SU(2)$_L$ gauge coupling constant $g_L$ and $M$ 
the $W^\pm$ mass $M_W$, with the tree-level relation to Fermi's constant, 
$G_F/\sqrt{2}=g^2_L/(8M^2_W)$.

A complex value for either of the coupling coefficients 
$\left\{S,V,T\right\}_{1,2,3,4}^{\eta_a,\eta_b}$ leads to
CP violation. The indices $\eta_a$ and $\eta_b$
correspond to the neutrino helicities $\eta_a$ or $\eta_b$,
while the lepton helicities are then identical or opposity depending
on the chiral structure of the coupling operator.
The tensor couplings $T^{\eta_a,\eta_b}_1$ and $T^{\eta_a,\eta_b}_4$
contribute only if $\eta_a=-\eta_b$, while the couplings
$T^{\eta_a,\eta_b}_2$ and $T^{\eta_a,\eta_b}_3$ contribute only if
$\eta_a=\eta_b$. 

For Majorana neutrinos, the parametrization is
\begin{equation}
{\cal L}_{\rm eff}=4\frac{g^2}{8M^2}\left[{\cal L}_M+{\cal L}^\dagger_M\right]
\ ,
\end{equation}
where
\begin{equation}
\begin{array}{r l}
{\cal L}_M=S^{\eta_a,\eta_b}\overline{\nu_a}P_{-\eta_a}\ell_i\
\overline{\ell_j}P_{\eta_b}\nu_b\ &+\
V^{\eta_a,\eta_b}\overline{\nu_a}\gamma^\mu P_{\eta_a}\ell_i\
\overline{\ell_j}\gamma_\mu P_{\eta_b}\nu_b\ \\
&+\
\frac{1}{2}T^{\eta_a,\eta_b}\overline{\nu_a}\sigma^{\mu\nu}P_{-\eta_a}\ell_i\
\overline{\ell_j}\sigma_{\mu\nu}P_{\eta_b}\nu_b\ .
\end{array}
\label{eq:Maj}
\end{equation}
Compared to the definitions above, and given the property 
$\psi^c_M=\psi_M$, the correspondence between the effective coupling 
coefficients in the Majorana and the Dirac cases is obvious.
Note that in the Dirac case, the total neutrino number is conserved only
for couplings of type 1 and 4, $\{S,V,T\}^{\eta_a,\eta_b}_{1,4}$,
whereas the couplings of type 2 and 3, $\{S,V,T\}^{\eta_a,\eta_b}_{2,3}$, 
violate that quantum number by two units.

In the electroweak Standard Model, besides the normalization
factor $4g^2/(8M^2)=4g^2_L/(8M^2_W)$, in order to identify the nonvanishing
couplings, different situations must be 
distinguished depending on whether only $W^\pm$ or only $Z_0$ exchanges 
are involved, or both. 

Purely $W^\pm$ exchange processes arise when $a=i$, $b=j$, $a\ne b$ 
and $i\ne j$, in which case the only nonvanishing effectif coupling is
the pure $(V-A)$ one, $V^{-,-}_1=-1$. For purely $Z_0$ neutral current 
processes which arise when $(a=b)\ne(i=j)$, 
the only nonvanishing couplings are $S^{-,-}_1=\sin^2\theta_W$ and
$V^{-,-}_1=\frac{1}{4}\left(1-2\sin^2\theta_W\right)$,
$\theta_W$ being the electroweak gauge mixing angle.
Finally, charged as well as neutral exchanges both contribute only when
all four fermion flavours are identical,
($a=b=i=j$), leading to the only nonvanishing couplings
$S^{-,-}_1=\sin^2\theta_W$ and 
$V^{-,-}_1=\frac{1}{4}\left(-1-2\sin^2\theta_W\right)$.
In the last two situations, ${\cal L}_D$ and ${\cal L}^\dagger_D$, 
or ${\cal L}_M$ and ${\cal L}^\dagger_M$, are identical.
Any extra coupling beyond these ones thus corresponds to some new physics
beyond the SM.

The remainder of the calculation proceeds straightforwardly. Given
any choice of external states for the in-coming and out-going particles
including their helicities, the substitution of the effective
Lagrangian operator enables the direct evaluation of the
matrix element ${\cal M}$ using the Fock algebra of the creation and
annihilation operators. Rather than computing $|{\cal M}|^2$ through the 
usual trace techniques, it is far more efficient to substitute for the 
$u(\vec{k},\eta)$ and $v(\vec{k},\eta)$ spinors. One then readily
obtains the value for ${\cal M}$ as a function of $\theta$. 

\section{Neutrino Pair Annihilation}
\label{Sect4}

In the Dirac case, neutrino pair annihilations are labelled as,

\begin{center}
\noindent\underline{$(ab)(ij)$ Dirac neutrino annihilations}\\
\begin{tabular}{r c l}
ab1: $\nu_a+\nu_b\rightarrow\ell^-_i+\ell^+_j$&\ \ \ ,\ \ \ &
ab2: $\nu_a+\nu_b\rightarrow\ell^+_i+\ell^-_j$\ \ \ ,\\
ab3: $\nu_a+\bar{\nu}_b\rightarrow\ell^-_i+\ell^+_j$&\ \ \ ,\ \ \ &
ab4: $\nu_a+\bar{\nu}_b\rightarrow\ell^+_i+\ell^-_j$\ \ \ ,\\
ab5: $\bar{\nu}_a+\nu_b\rightarrow\ell^-_i+\ell^+_j$&\ \ \ ,\ \ \ &
ab6: $\bar{\nu}_a+\nu_b\rightarrow\ell^+_i+\ell^-_j$\ \ \ ,\\
ab7: $\bar{\nu}_a+\bar{\nu}_b\rightarrow\ell^-_i+\ell^+_j$&\ \ \ ,\ \ \ &
ab8: $\bar{\nu}_a+\bar{\nu}_b\rightarrow\ell^+_i+\ell^-_j$\ \ \ ,\\
\end{tabular}
\end{center}
\noindent while in the Majorana case,

\begin{center}
\noindent\underline{$(ab)(ij)$ Majorana neutrino annihilations}

Mab1: $\nu_a+\nu_b\rightarrow\ell^-_i+\ell^+_j$\ \ \ ,\ \ \ 
Mab2: $\nu_a+\nu_b\rightarrow\ell^+_i+\ell^-_j$\ .\\
\end{center}

Due to common angular-momentum selection rules, 
the matrix element ${\cal M}$ for all these ten processes is
\begin{equation}
\begin{array}{l l}
{\cal M}_{(ab)(ij)}=&\\
 & \\
=-4s\left(\frac{g^2}{8M^2}\right)\,&N_1\,\delta_{ij} 
\left\{\delta_{ab}\delta_{\eta_i}^{-\eta_a}\delta_{\eta_j}^{-\eta_b}
\left[A_{11}\sin^2\theta/2+2\delta_{\eta_a,\eta_b}B_{11}(1+\cos^2\theta/2)
\right]\right.\\
 & + \delta_{ab}\delta_{\eta_i}^{\eta_a}\delta_{\eta_j}^{\eta_b}
C_{11}\left[(1+\eta_a\eta_b)-(1-\eta_a\eta_b)\cos^2\theta/2\right] \\
 & + \delta_{\eta_i}^{-\eta_b}\delta_{\eta_j}^{-\eta_a}\eta_a\eta_bD_1
\left[A_{12}\cos^2\theta/2+2\delta_{\eta_a,\eta_b}B_{12}(1+\sin^2\theta/2)
\right] \\
 & +
\left.\delta_{\eta_i}^{\eta_b}\delta_{\eta_j}^{\eta_a}\eta_a\eta_b D_1
C_{12}\left[(1+\eta_a\eta_b)-(1-\eta_a\eta_b)\sin^2\theta/2\right]\right\}\\
\ \ \ -4s\left(\frac{g^2}{8M^2}\right)\,&N_2\,
\left\{\delta_{ab}\delta_{\eta_i}^{-\eta_b}\delta_{\eta_j}^{-\eta_a}
\left[A_{21}\cos^2\theta/2+2\delta_{\eta_a,\eta_b}B_{21}(1+\sin^2\theta/2)
\right]\right.\\
 & +
\delta_{ab}\delta_{\eta_i}^{\eta_b}\delta_{\eta_j}^{\eta_a}
C_{21}\left[(1+\eta_a\eta_b)-(1-\eta_a\eta_b)\sin^2\theta/2\right] \\
 & +
\delta_{\eta_i}^{-\eta_a}\delta_{\eta_j}^{-\eta_b}\eta_a\eta_bD_2
\left[A_{22}\sin^2\theta/2+2\delta_{\eta_a,\eta_b}B_{22}(1+\cos^2\theta/2)
\right] \\
 & +
\left.\delta_{\eta_i}^{\eta_a}\delta_{\eta_j}^{\eta_b}\eta_a\eta_b D_2
C_{22}\left[(1+\eta_a\eta_b)-(1-\eta_a\eta_b)\cos^2\theta/2\right]\right\}\ ,
\end{array}
\label{eq:abij}
\end{equation}
where $\theta$ is the scattering angle between the neutrino of
flavour $a$ and the charged lepton of flavour $i$.
The particle helicities are $\eta_a$, $\eta_b$, $\eta_i$ and $\eta_j$,
respectively. Tables~\ref{Table:abij1} and \ref{Table:abij2}
list the values for the constant
phase factors $N_{1,2}$ and $D_{1,2}$ and the subsets of the scalar,
tensor and vector effective couplings constants, in that order,
defining the quantities $A_{11,12,21,22}$, $B_{11,12,21,22}$ and 
$C_{11,12,21,22}$, whether in the case of Dirac or Majorana neutrinos.

\section{Neutrino Pair Production}
\label{Sect5}

For the sake of completeness, neutrino pair production has also been
con\-si\-der\-ed. In the Dirac case, the following list applies,

\begin{center}
\noindent\underline{$(ij)(ab)$ Dirac processes}\\
\begin{tabular}{r c l}
ij1: $\ell^-_i+\ell^+_j\rightarrow \nu_a+\nu_b$&\ \ \ ,\ \ \ &
ij2: $\ell^+_i+\ell^-_j\rightarrow \nu_a+\nu_b$\ \ \ ,\\
ij3: $\ell^-_i+\ell^+_j\rightarrow \nu_a+\bar{\nu}_b$&\ \ \ ,\ \ \ &
ij4: $\ell^+_i+\ell^-_j\rightarrow \nu_a+\bar{\nu}_b$\ \ \ ,\\
ij5: $\ell^-_i+\ell^+_j\rightarrow \bar{\nu}_a+\nu_b$&\ \ \ ,\ \ \ &
ij6: $\ell^+_i+\ell^-_j\rightarrow \bar{\nu}_a+\nu_b$\ \ \ ,\\
ij7: $\ell^-_i+\ell^+_j\rightarrow \bar{\nu}_a+\bar{\nu}_b$&\ \ \ ,\ \ \ &
ij8: $\ell^+_i+\ell^-_j\rightarrow \bar{\nu}_a+\bar{\nu}_b$\ \ \ ,\\
\end{tabular}
\end{center}
\noindent while in the Majorana case

\begin{center}
\noindent\underline{$(ij)(ab)$ Majorana processes}

Mij1: $\ell^-_i+\ell^+_j\rightarrow \nu_a+\nu_b$\ \ \ ,\ \ \ 
Mij2: $\ell^+_i+\ell^-_j\rightarrow \nu_a+\nu_b$\ .\\
\end{center}

For all these ten processes, the amplitude ${\cal M}$ is of the form
\begin{equation}
\begin{array}{l l}
{\cal M}_{(ij)(ab)}=\\
 & \\
=4s\left(\frac{g^2}{8M^2}\right)\,&N_1\,\delta_{ij} 
\left\{\delta_{ab}\delta_{\eta_i}^{-\eta_a}\delta_{\eta_j}^{-\eta_b}
\left[A_{11}\sin^2\theta/2+2\delta_{\eta_a,\eta_b}B_{11}(1+\cos^2\theta/2)
\right] \right.\\
 & -
\delta_{ab}\delta_{\eta_i}^{\eta_a}\delta_{\eta_j}^{\eta_b}C_{11}
\left[(1+\eta_a\eta_b)-(1-\eta_a\eta_b)\cos^2\theta/2\right] \\
 & +
\delta_{\eta_i}^{-\eta_b}\delta_{\eta_j}^{-\eta_a} D_1
\left[A_{12}\cos^2\theta/2+2\delta_{\eta_a,\eta_b}B_{12}(1+\sin^2\theta/2)
\right]\\
 & -
\left.\delta_{\eta_i}^{\eta_b}\delta_{\eta_j}^{\eta_a} D_1 C_{12}
\left[(1+\eta_a\eta_b)-(1-\eta_a\eta_b)\sin^2\theta/2\right]\right\} \\
\ +4s\left(\frac{g^2}{8M^2}\right)\,&N_2\,
\left\{\delta_{ab}\delta_{\eta_i}^{-\eta_b}\delta_{\eta_j}^{-\eta_a}
\left[A_{21}\cos^2\theta/2+2\delta_{\eta_a,\eta_b}B_{21}(1+\sin^2\theta/2)
\right] \right.\\
 & -
\delta_{ab}\delta_{\eta_i}^{\eta_b}\delta_{\eta_j}^{\eta_a} C_{21}
\left[(1+\eta_a\eta_b)-(1-\eta_a\eta_b)\sin^2\theta/2\right] \\
 & +
\delta_{\eta_i}^{-\eta_a}\delta_{\eta_j}^{-\eta_b} D_2
\left[A_{22}\sin^2\theta/2+2\delta_{\eta_a,\eta_b}B_{22}(1+\cos^2\theta/2)
\right] \\
 & -
\left.\delta_{\eta_i}^{\eta_a}\delta_{\eta_j}^{\eta_b} D_2 C_{22}
\left[(1+\eta_a\eta_b)-(1-\eta_a\eta_b)\cos^2\theta/2\right]\right\}\ ,
\end{array}
\label{eq:ijab}
\end{equation}
$\theta$ being the angle between the lepton of flavour $i$ and the
neutrino of flavour $a$. The different factors and coefficients appearing
in this expression are detailed in Tables~\ref{Table:ijab1} and
\ref{Table:ijab2}, whether in the case of Dirac or Majorana neutrinos.

\section{Neutrino Scattering}
\label{Sect6}

In the case of neutrino scattering onto a 
charged lepton, we list the results only for the classes of processes
$(ai)(bj)$ and $(aj)(bi)$, since the other two classes $(bi)(aj)$ and
$(bj)(ai)$ may be obtained by appropriate permutations.\cite{Mendy1,Mendy2}

\subsection{$(ai)(bj)$ neutrino scattering processes}
\label{Subsect6.1}

In the case of Dirac neutrinos, the list of processes is

\begin{center}
\noindent\underline{$(ai)(bj)$ Dirac processes}\\
\begin{tabular}{r c l}
ai1: $\nu_a+\ell^-_i\rightarrow \nu_b+\ell^-_j$&\ \ \ ,\ \ \ &
ai2: $\nu_a+\ell^+_i\rightarrow \nu_b+\ell^+_j$\ \ \ ,\\
ai3: $\nu_a+\ell^-_i\rightarrow \bar{\nu}_b+\ell^-_j$&\ \ \ ,\ \ \ &
ai4: $\nu_a+\ell^+_i\rightarrow \bar{\nu}_b+\ell^+_j$\ \ \ ,\\
ai5: $\bar{\nu}_a+\ell^-_i\rightarrow \nu_b+\ell^-_j$&\ \ \ ,\ \ \ &
ai6: $\bar{\nu}_a+\ell^+_i\rightarrow \nu_b+\ell^+_j$\ \ \ ,\\
ai7: $\bar{\nu}_a+\ell^-_i\rightarrow \bar{\nu}_b+\ell^-_j$&\ \ \ ,\ \ \ &
ai8: $\bar{\nu}_a+\ell^+_i\rightarrow \bar{\nu}_b+\ell^+_j$\ \ \ ,\\
\end{tabular}
\end{center}
\noindent while in the Majorana case

\begin{center}
\noindent\underline{$(ai)(bj)$ Majorana processes}

Mai1: $\nu_a+\ell^-_i\rightarrow \nu_b+\ell^-_j$\ \ \ ,\ \ \ 
Mai2: $\nu_a+\ell^+_i\rightarrow \nu_b+\ell^+_j$\ .\\
\end{center}

The general amplitude ${\cal M}$ then reads in all ten cases
\begin{equation}
\begin{array}{l l}
{\cal M}_{(ai)(bj)}=&\\
& \\
=4s\left(\frac{g^2}{8M^2}\right)\,&N_1\,\delta_{ij} 
\left\{\delta_{ab}\delta_{\eta_i}^{\eta_a}\delta_{\eta_j}^{\eta_b}
\left[A_{11}-2\delta_{\eta_a,-\eta_b}B_{11}(\cos^2\theta/2-\sin^2\theta/2)
\right]\right.\\
 & +
\delta_{ab}\delta_{\eta_i}^{-\eta_a}\delta_{\eta_j}^{-\eta_b}
C_{11}\left[1+\eta_a\eta_b(\cos^2\theta/2-\sin^2\theta/2)\right] \\
 & +
\delta_{\eta_i}^{-\eta_b}\delta_{\eta_j}^{-\eta_a}\eta_a\eta_b D_1
\left[A_{12}\cos^2\theta/2-2\delta_{\eta_a,-\eta_b}B_{12}(1+\sin^2\theta/2)
\right]\\
 & +
\left.\delta_{\eta_i}^{\eta_b}\delta_{\eta_j}^{\eta_a}\eta_a\eta_b D_1
C_{12}\left[(1+\eta_a\eta_b)-(1-\eta_a\eta_b)\sin^2\theta/2)\right]\right\} \\
\ +4s\left(\frac{g^2}{8M^2}\right)\,&N_2\,
\left\{\delta_{ab}\delta_{\eta_i}^{-\eta_b}\delta_{\eta_j}^{-\eta_a}
\left[A_{21}\cos^2\theta/2-2\delta_{\eta_a,-\eta_b}B_{21}(1+\sin^2\theta/2)
\right]\right.\\
 & +
\delta_{ab}\delta_{\eta_i}^{\eta_b}\delta_{\eta_j}^{\eta_a}
C_{21}\left[(1+\eta_a\eta_b)-(1-\eta_a\eta_b)\sin^2\theta/2\right] \\
 & +
\delta_{\eta_i}^{\eta_a}\delta_{\eta_j}^{\eta_b}\eta_a\eta_b D_2 
\left[A_{22}-2\delta_{\eta_a,-\eta_b}B_{22}(\cos^2\theta/2-\sin^2\theta/2)
\right]\\
 & +
\left.\delta_{\eta_i}^{-\eta_a}\delta_{\eta_j}^{-\eta_b}\eta_a\eta_b D_2
C_{22}\left[1+\eta_a\eta_b(\cos^2\theta/2-\sin^2\theta/2)\right]\right\}\ ,
\end{array}
\label{eq:aibj}
\end{equation}
$\theta$ being the neutrino scattering angle. The list of factors and
coefficients appearing in this expression is detailed in
Tables~\ref{Table:aibj1} and \ref{Table:aibj2}, both in the Dirac and 
in the Majorana case.

\subsection{$(aj)(bi)$ neutrino scattering processes}
\label{Subsect6.2}

The list of processes in the Dirac case is

\begin{center}
\noindent\underline{$(aj)(bi)$ Dirac processes}\\
\begin{tabular}{r c l}
aj1: $\nu_a+\ell^-_j\rightarrow \nu_b+\ell^-_i$&\ \ \ ,\ \ \ &
aj2: $\nu_a+\ell^+_j\rightarrow \nu_b+\ell^+_i$\ \ \ ,\\
aj3: $\nu_a+\ell^-_j\rightarrow \bar{\nu}_b+\ell^-_i$&\ \ \ ,\ \ \ &
aj4: $\nu_a+\ell^+_j\rightarrow \bar{\nu}_b+\ell^+_i$\ \ \ ,\\
aj5: $\bar{\nu}_a+\ell^-_j\rightarrow \nu_b+\ell^-_i$&\ \ \ ,\ \ \ &
aj6: $\bar{\nu}_a+\ell^+_j\rightarrow \nu_b+\ell^+_i$\ \ \ ,\\
aj7: $\bar{\nu}_a+\ell^-_j\rightarrow \bar{\nu}_b+\ell^-_i$&\ \ \ ,\ \ \ &
aj8: $\bar{\nu}_a+\ell^+_j\rightarrow \bar{\nu}_b+\ell^+_i$\ \ \ ,\\
\end{tabular}
\end{center}

\noindent while in the Majorana case

\begin{center}
\noindent\underline{$(aj)(bi)$ Majorana processes}

Maj1: $\nu_a+\ell^-_j\rightarrow \nu_b+\ell^-_i$\ \ \ ,\ \ \ 
Maj2: $\nu_a+\ell^+_j\rightarrow \nu_b+\ell^+_i$\ .\\
\end{center}

The general scattering amplitude ${\cal M}$ is of the form
\begin{equation}
\begin{array}{l l}
{\cal M}_{(aj)(bi)}=&\\
 & \\
=4s\left(\frac{g^2}{8M^2}\right)\,&N_1\,\delta_{ij}
\left\{\delta_{ab}\delta_{\eta_i}^{-\eta_a}\delta_{\eta_j}^{-\eta_b}
\left[A_{11}\cos^2\theta/2-2\delta_{\eta_a,-\eta_b}B_{11}(1+\sin^2\theta/2)
\right]\right.\\
 & +
\delta_{ab}\delta_{\eta_i}^{\eta_a}\delta_{\eta_j}^{\eta_b}
C_{11}\left[(1+\eta_a\eta_b)-(1-\eta_a\eta_b)\sin^2\theta/2\right] \\
 & +
\delta_{\eta_i}^{\eta_b}\delta_{\eta_j}^{\eta_a}\eta_a\eta_b D_1
\left[A_{12}-2\delta_{\eta_a,-\eta_b}B_{12}(\cos^2\theta/2-\sin^2\theta/2)
\right]\\
 & +
\left.\delta_{\eta_i}^{-\eta_b}\delta_{\eta_j}^{-\eta_a}\eta_a\eta_b D_1
C_{12}\left[1+\eta_a\eta_b(\cos^2\theta/2-\sin^2\theta/2)\right]\right\} \\
\ +4s\left(\frac{g^2}{8M^2}\right)\,&N_2\, 
\left\{\delta_{ab}\delta_{\eta_i}^{\eta_b}\delta_{\eta_j}^{\eta_a}
\left[A_{21}-2\delta_{\eta_a,-\eta_b}B_{21}(\cos^2\theta/2-\sin^2\theta/2)
\right]\right.\\
 & +
\delta_{ab}\delta_{\eta_i}^{-\eta_b}\delta_{\eta_j}^{-\eta_a}
C_{21}\left[1+\eta_a\eta_b(\cos^2\theta/2-\sin^2\theta/2)\right] \\
 & +
\delta_{\eta_i}^{-\eta_a}\delta_{\eta_j}^{-\eta_b}\eta_a\eta_b D_2 
\left[A_{22}\cos^2\theta/2-2\delta_{\eta_a,-\eta_b}B_{22}(1+\sin^2\theta/2)
\right]\\
 & +
\left.\delta_{\eta_i}^{\eta_a}\delta_{\eta_j}^{\eta_b}\eta_a\eta_b D_2
C_{22}\left[(1+\eta_a\eta_b)-(1-\eta_a\eta_b)\sin^2\theta/2)\right]\right\}\ ,
\end{array}
\label{eq:ajbi}
\end{equation}
the angle $\theta$ being that of the scattered neutrino. 
Tables~\ref{Table:ajbi1} and \ref{Table:ajbi2}
list the relevant factors and coefficients both in the Dirac and in the
Majorana case.

\section{Concluding Remarks}
\label{Sect7}

The above general results\cite{Mendy1,Mendy2} provide means to assess 
directly the sensitivity of neutrino beam experiments in the energy range
up to a few ten's of GeV's to different fundamental issues of physics in 
the neutrino sector, whether new interactions beyond the Standard Model, 
whether the Dirac or Majorana character of neutrinos. By lack
of space, only one illustration of the latter instance is presented.

Scalar or tensor couplings being typically less well constrained than 
vector ones, let us consider an extra scalar
interaction, for either of the following two elastic scattering reactions,
\begin{equation}
\nu_\mu+e^-\rightarrow\nu_\mu+e^-\ \ \ ,\ \ \ 
\nu_\mu+\mu^-\rightarrow\nu_\mu+\mu^-\ .
\end{equation}
These reactions are of the ``ai1" type in the $(ai)(bj)$ class,
with $a=b\ne i=j$ in the first case, and $a=b=i=j$ in the second.
Assuming that beyond the couplings of the SM, $S^{+,-}_1$ is the
sole nonvanishing extra interaction, and considering
an unpolarized measurement, the sum over all polarization states reads,
\begin{equation}
\begin{array}{r l}
\sum_{\rm pol.}&|{\cal M}|^2=(4s)^2\left(\frac{g^2}{8M^2}\right)^2\times\\
&\times
\left\{\left[4{\rm Re}\,V^{-,-}_1\right]^2+
\left[{\rm Re}\,S^{-,-}_1\right]^2(1+\cos\theta)^2+
\frac{1}{4}|S^{+,-}_1|^2(1\pm\cos\theta)^2\right\}\ ,
\end{array}
\end{equation}
where in the last term the upper sign corresponds to the Dirac case,
and the lower sign to the Majorana case. Hence indeed,
any interaction whose chirality structure differs from the SM one
leads to processes in which the angular dependency discriminates
between Dirac and Majorana neutrinos. Taking as an illustration a value
$|S^{+,-}_1|=0.10$ which is a typical upper-bound on such a coupling in 
the leptonic $(e\mu)$ sector\cite{PDG}, one finds a 10\% sensitivity in 
the forward-backward asymmetry, certainly a possibility worth to be explored
further within the context of realistic foreseen experimental conditions. 
In other words, neutrino factories may offer an alternative to
neutrinoless double $\beta$-decay\cite{Klapdor} in establishing the Dirac
or Majorana character of neutrinos.

The main purpose of this work\cite{Mendy1,Mendy2} has been to provide 
the general results for the Feynman amplitudes for all possible 
$2\rightarrow 2$ processes with two neutrinos. On that basis, it should now 
be possible to develop a detailed and dedicated analysis, extending similar 
work within restricted classes of effective couplings,\cite{Valle} 
of the potential reach of different such reactions 
towards the above physics issues, inclusive of the possible discrimination
between Dirac and Majorana neutrinos, given a specific design both of neutrino 
beams and their intensities, and of detector set-ups. Besides the great 
interest to be found in neutrino scattering experiments, the possibilities 
offered by intersecting neutrino beams should also not be dismissed offhand 
without first a dedicated assessment as well, the more so since they could 
possibly run in pa\-ra\-si\-tic mode in conjunction with other experiments 
given a proper geometry for the neutrino beams.

\section*{\bf Acknowledgements}

J.E.B.M. acknowledges the financial support of the ``Coop\'eration
Universitaire au D\'evelop\-pe\-ment, Commission Interuniversitaire
Francophone" (CUD-CIUF) of the Belgian French Speaking Community which
enabled him to pursue his Ph.~D. at the
``Institut de Math\'ematiques et de Sciences Physiques (IMSP)" (Benin),
and wishes to thank 
the Institute of Nuclear Physics (Catholic University
of Louvain, Belgium) for its hos\-pi\-ta\-li\-ty while this work was being
pursued. J.G. wishes to thank the C.N.~Yang Institute for Theoretical Physics 
(State University of New York at Stony Brook, USA) for its hos\-pi\-ta\-li\-ty
during the Summer 2001 while part of this work was completed.

\clearpage

\begin{table}
\caption[]{The constant factors appearing in (\ref{eq:abij}) for the
first five $(ab)(ij)$ Dirac neutrino annihilation processes.}
\begin{center}
\begin{tabular}{|c||c|c|c|c|c|}
\hline
 & ab1 & ab2 & ab3 & ab4 & ab5 \\
\hline\hline
$N_1$ & 
$\eta_a\lambda_a^*$ & $\eta_a\lambda_b^*$ & 
$\eta_a$ & $\eta_a$ & $\eta_a$ \\
\hline\hline
$A_{11}$ &
$S_2^{\eta_b,\eta_a}$ & $S_3^{\eta_b,\eta_a*}$ &
$S_1^{-\eta_b,\eta_a}$ & $S_4^{-\eta_b,\eta_a*}$ &
$S_4^{\eta_b,-\eta_a}$ \\
\hline
$B_{11}$ &
$T_2^{\eta_b,\eta_a}$ & $T_3^{\eta_b,\eta_a*}$ &
$T_1^{-\eta_b,\eta_a}$ & $T_4^{-\eta_b,\eta_a*}$ &
$T_4^{\eta_b,-\eta_a}$ \\
\hline
$C_{11}$ &
$V_2^{\eta_b,\eta_a}$ & $V_3^{\eta_b,\eta_a*}$ &
$V_1^{-\eta_b,\eta_a}$ & $V_4^{-\eta_b,\eta_a*}$ &
$V_4^{\eta_b,-\eta_a}$ \\
\hline\hline
$D_1$ &
$1$ & $1$ & $\lambda_a^*\lambda_b$ & $1$ & $1$ \\
\hline
$A_{12}$ &
$S_2^{\eta_a,\eta_b}$ & $S_3^{\eta_a,\eta_b*}$ &
$S_4^{\eta_a,-\eta_b}$ & $S_1^{\eta_a,-\eta_b*}$ &
$S_1^{-\eta_a,\eta_b}$ \\
\hline
$B_{12}$ &
$T_2^{\eta_a,\eta_b}$ & $T_3^{\eta_a,\eta_b*}$ &
$T_4^{\eta_a,-\eta_b}$ & $T_1^{\eta_a,-\eta_b*}$ &
$T_1^{-\eta_a,\eta_b}$ \\
\hline
$C_{12}$ &
$V_2^{\eta_a,\eta_b}$ & $V_3^{\eta_a,\eta_b*}$ &
$V_4^{\eta_a,-\eta_b}$ & $V_1^{\eta_a,-\eta_b*}$ &
$V_1^{-\eta_a,\eta_b}$ \\
\hline\hline\hline
$N_2$ & 
$\eta_b\lambda_b^*$ & $\eta_b\lambda_a^*$ & 
$\eta_b$ & $\eta_b$ & $\eta_b$ \\
\hline\hline
$A_{21}$ &
$S_3^{\eta_b,\eta_a*}$ & $S_2^{\eta_b,\eta_a}$ &
$S_4^{-\eta_b,\eta_a*}$ & $S_1^{-\eta_b,\eta_a}$ &
$S_1^{\eta_b,-\eta_a*}$ \\
\hline
$B_{21}$ &
$T_3^{\eta_b,\eta_a*}$ & $T_2^{\eta_b,\eta_a}$ &
$T_4^{-\eta_b,\eta_a*}$ & $T_1^{-\eta_b,\eta_a}$ &
$T_1^{\eta_b,-\eta_a*}$ \\
\hline
$C_{21}$ &
$V_3^{\eta_b,\eta_a*}$ & $V_2^{\eta_b,\eta_a}$ &
$V_4^{-\eta_b,\eta_a*}$ & $V_1^{-\eta_b,\eta_a}$ &
$V_1^{\eta_b,-\eta_a*}$ \\
\hline\hline
$D_2$ &
$1$ & $1$ & $1$ & $\lambda_a^*\lambda_b$ & $\lambda_a\lambda_b^*$ \\
\hline
$A_{22}$ &
$S_3^{\eta_a,\eta_b*}$ & $S_2^{\eta_a,\eta_b}$ &
$S_1^{\eta_a,-\eta_b*}$ & $S_4^{\eta_a,-\eta_b}$ &
$S_4^{-\eta_a,\eta_b*}$ \\
\hline
$B_{22}$ &
$T_3^{\eta_a,\eta_b*}$ & $T_2^{\eta_a,\eta_b}$ &
$T_1^{\eta_a,-\eta_b*}$ & $T_4^{\eta_a,-\eta_b}$ &
$T_4^{-\eta_a,\eta_b*}$ \\
\hline
$C_{22}$ &
$V_3^{\eta_a,\eta_b*}$ & $V_2^{\eta_a,\eta_b}$ &
$V_1^{\eta_a,-\eta_b*}$ & $V_4^{\eta_a,-\eta_b}$ &
$V_4^{-\eta_a,\eta_b*}$ \\
\hline
\end{tabular}
\label{Table:abij1}
\end{center}
\end{table}

\begin{table}
\caption[]{The constant factors appearing in (\ref{eq:abij}) for the
last three $(ab)(ij)$ Dirac neutrino annihilation processes, and the
two Majorana neutrino ones.}
\begin{center}
\begin{tabular}{|c||c|c|c||c|c|}
\hline
& ab6 & ab7 & ab8 & Mab1 & Mab2 \\
\hline\hline
$N_1$ & 
$\eta_a$ & $\eta_a\lambda_b$ & $\eta_a\lambda_a$ & 
$\eta_a\lambda_a^*$ & $\eta_a\lambda_b^*$ \\
\hline\hline
$A_{11}$ &
$S_1^{\eta_b,-\eta_a*}$ &
$S_3^{-\eta_b,-\eta_a}$ & $S_2^{-\eta_b,-\eta_a*}$ &
$S^{-\eta_b,\eta_a}$ & $S^{\eta_b,-\eta_a*}$ \\
\hline
$B_{11}$ &
$T_1^{\eta_b,-\eta_a*}$ &
$T_3^{-\eta_b,-\eta_a}$ & $T_2^{-\eta_b,-\eta_a*}$ &
$T^{-\eta_b,\eta_a}$ & $T^{\eta_b,-\eta_a*}$ \\
\hline
$C_{11}$ &
$V_1^{\eta_b,-\eta_a*}$ &
$V_3^{-\eta_b,-\eta_a}$ & $V_2^{-\eta_b,-\eta_a*}$ &
$V^{-\eta_b,\eta_a}$ & $V^{\eta_b,-\eta_a*}$ \\
\hline\hline
$D_1$ &
$\lambda_a\lambda_b^*$ & $1$ & $1$ & $1$ & $1$ \\
\hline
$A_{12}$ &
$S_4^{-\eta_a,\eta_b*}$ &
$S_3^{-\eta_a,-\eta_b}$ & $S_2^{-\eta_a,-\eta_b*}$ &
$S^{-\eta_a,\eta_b}$ & $S^{\eta_a,-\eta_b*}$ \\
\hline
$B_{12}$ &
$T_4^{-\eta_a,\eta_b*}$ &
$T_3^{-\eta_a,-\eta_b}$ & $T_2^{-\eta_a,-\eta_b*}$ &
$T^{-\eta_a,\eta_b}$ & $T^{\eta_a,-\eta_b*}$ \\
\hline
$C_{12}$ &
$V_4^{-\eta_a,\eta_b*}$ &
$V_3^{-\eta_a,-\eta_b}$ & $V_2^{-\eta_a,-\eta_b*}$ &
$V^{-\eta_a,\eta_b}$ & $V^{\eta_a,-\eta_b*}$ \\
\hline\hline\hline
$N_2$ & 
$\eta_b$ & $\eta_b\lambda_a$ & 
$\eta_b\lambda_b$ & $\eta_b\lambda_b^*$ & $\eta_b\lambda_a^*$ \\
\hline\hline
$A_{21}$ &
$S_4^{\eta_b,-\eta_a}$ &
$S_2^{-\eta_b,-\eta_a*}$ & $S_3^{-\eta_b,-\eta_a}$ &
$S^{\eta_b,-\eta_a*}$ & $S^{-\eta_b,\eta_a}$ \\
\hline
$B_{21}$ &
$T_4^{\eta_b,-\eta_a}$ &
$T_2^{-\eta_b,-\eta_a*}$ & $T_3^{-\eta_b,-\eta_a}$ &
$T^{\eta_b,-\eta_a*}$ & $T^{-\eta_b,\eta_a}$ \\
\hline
$C_{21}$ &
$V_4^{\eta_b,-\eta_a}$ &
$V_2^{-\eta_b,-\eta_a*}$ & $V_3^{-\eta_b,-\eta_a}$ &
$V^{\eta_b,-\eta_a*}$ & $V^{-\eta_b,\eta_a}$ \\
\hline\hline
$D_2$ &
$1$ & $1$ & $1$ & $1$ & $1$ \\
\hline
$A_{22}$ &
$S_1^{-\eta_a,\eta_b}$ &
$S_2^{-\eta_a,-\eta_b*}$ & $S_3^{-\eta_a,-\eta_b}$ &
$S^{\eta_a,-\eta_b*}$ & $S^{-\eta_a,\eta_b}$ \\
\hline
$B_{22}$ &
$T_1^{-\eta_a,\eta_b}$ &
$T_2^{-\eta_a,-\eta_b*}$ & $T_3^{-\eta_a,-\eta_b}$ &
$T^{\eta_a,-\eta_b*}$ & $T^{-\eta_a,\eta_b}$ \\
\hline
$C_{22}$ &
$V_1^{-\eta_a,\eta_b}$ &
$V_2^{-\eta_a,-\eta_b*}$ & $V_3^{-\eta_a,-\eta_b}$ &
$V^{\eta_a,-\eta_b*}$ & $V^{-\eta_a,\eta_b}$ \\
\hline
\end{tabular}
\label{Table:abij2}
\end{center}
\end{table}

\vspace{10pt}

\begin{table}
\caption[]{The constant factors appearing in (\ref{eq:ijab}) for the
first five $(ij)(ab)$ Dirac neutrino pair production processes.}
\begin{center}
\begin{tabular}{|c||c|c|c|c|c|}
\hline
 & ij1 & ij2 & ij3 & ij4 & ij5 \\
\hline\hline
$N_1$ & 
$\eta_a\lambda_a$ & $\eta_a\lambda_b$ & 
$\eta_a$ & $\eta_a$ & $\eta_a$ \\
\hline\hline
$A_{11}$ &
$S_2^{\eta_b,\eta_a*}$ & $S_3^{\eta_b,\eta_a}$ &
$S_1^{-\eta_b,\eta_a*}$ & $S_4^{-\eta_b,\eta_a}$ &
$S_4^{\eta_b,-\eta_a*}$ \\ 
\hline
$B_{11}$ &
$T_2^{\eta_b,\eta_a*}$ & $T_3^{\eta_b,\eta_a}$ &
$T_1^{-\eta_b,\eta_a*}$ & $T_4^{-\eta_b,\eta_a}$ &
$T_4^{\eta_b,-\eta_a*}$ \\ 
\hline
$C_{11}$ &
$V_2^{\eta_b,\eta_a*}$ & $V_3^{\eta_b,\eta_a}$ &
$V_1^{-\eta_b,\eta_a*}$ & $V_4^{-\eta_b,\eta_a}$ &
$V_4^{\eta_b,-\eta_a*}$ \\ 
\hline\hline
$D_1$ &
$1$ & $1$ & $\lambda_a\lambda_b^*$ & $1$ & $1$ \\
\hline
$A_{12}$ &
$S_2^{\eta_a,\eta_b*}$ & $S_3^{\eta_a,\eta_b}$ &
$S_4^{\eta_a,-\eta_b*}$ & $S_1^{\eta_a,-\eta_b}$ &
$S_1^{-\eta_a,\eta_b*}$ \\
\hline
$B_{12}$ &
$T_2^{\eta_a,\eta_b*}$ & $T_3^{\eta_a,\eta_b}$ &
$T_4^{\eta_a,-\eta_b*}$ & $T_1^{\eta_a,-\eta_b}$ &
$T_1^{-\eta_a,\eta_b*}$ \\
\hline
$C_{12}$ &
$V_2^{\eta_a,\eta_b*}$ & $V_3^{\eta_a,\eta_b}$ &
$V_4^{\eta_a,-\eta_b*}$ & $V_1^{\eta_a,-\eta_b}$ &
$V_1^{-\eta_a,\eta_b*}$ \\
\hline\hline\hline
$N_2$ & 
$\eta_a\lambda_b$ & $\eta_a\lambda_a$ & 
$\eta_a$ & $\eta_a$ & $\eta_a$ \\
\hline\hline
$A_{21}$ &
$S_3^{\eta_b,\eta_a}$ & $S_2^{\eta_b,\eta_a*}$ &
$S_4^{-\eta_b,\eta_a}$ & $S_1^{-\eta_b,\eta_a*}$ &
$S_1^{\eta_b,-\eta_a}$ \\ 
\hline
$B_{21}$ &
$T_3^{\eta_b,\eta_a}$ & $T_2^{\eta_b,\eta_a*}$ &
$T_4^{-\eta_b,\eta_a}$ & $T_1^{-\eta_b,\eta_a*}$ &
$T_1^{\eta_b,-\eta_a}$ \\
\hline
$C_{21}$ &
$V_3^{\eta_b,\eta_a}$ & $V_2^{\eta_b,\eta_a*}$ &
$V_4^{-\eta_b,\eta_a}$ & $V_1^{-\eta_b,\eta_a*}$ &
$V_1^{\eta_b,-\eta_a}$ \\
\hline\hline
$D_2$ &
$1$ & $1$ & $1$ & $\lambda_a\lambda_b^*$ & $\lambda_a^*\lambda_b$ \\
\hline
$A_{22}$ &
$S_3^{\eta_a,\eta_b}$ & $S_2^{\eta_a,\eta_b*}$ &
$S_1^{\eta_a,-\eta_b}$ & $S_4^{\eta_a,-\eta_b*}$ &
$S_4^{-\eta_a,\eta_b}$ \\
\hline
$B_{22}$ &
$T_3^{\eta_a,\eta_b}$ & $T_2^{\eta_a,\eta_b*}$ &
$T_1^{\eta_a,-\eta_b}$ & $T_4^{\eta_a,-\eta_b*}$ &
$T_4^{-\eta_a,\eta_b}$ \\
\hline
$C_{22}$ &
$V_3^{\eta_a,\eta_b}$ & $V_2^{\eta_a,\eta_b*}$ &
$V_1^{\eta_a,-\eta_b}$ & $V_4^{\eta_a,-\eta_b*}$ &
$V_4^{-\eta_a,\eta_b}$ \\
\hline
\end{tabular}
\label{Table:ijab1}
\end{center}
\end{table}

\begin{table}
\caption[]{The constant factors appearing in (\ref{eq:ijab}) for the
last three $(ij)(ab)$ Dirac neutrino pair production processes,
and the two Majorana neutrino ones.}
\begin{center}
\begin{tabular}{|c||c|c|c||c|c|}
\hline
 & ij6 & ij7 & ij8 & Mij1 & Mij2 \\
\hline\hline
$N_1$ & 
$\eta_a$ &
$\eta_a\lambda_b^*$ & $\eta_a\lambda_a^*$ & 
$\eta_a\lambda_a$ & $\eta_a\lambda_b$ \\
\hline\hline
$A_{11}$ &
$S_1^{\eta_b,-\eta_a}$ &
$S_3^{-\eta_b,-\eta_a*}$ & $S_2^{-\eta_b,-\eta_a}$ &
$S^{-\eta_b,\eta_a*}$ & $S^{\eta_b,-\eta_a}$ \\
\hline
$B_{11}$ &
$T_1^{\eta_b,-\eta_a}$ &
$T_3^{-\eta_b,-\eta_a*}$ & $T_2^{-\eta_b,-\eta_a}$ &
$T^{-\eta_b,\eta_a*}$ & $T^{\eta_b,-\eta_a}$ \\
\hline
$C_{11}$ &
$V_1^{\eta_b,-\eta_a}$ &
$V_3^{-\eta_b,-\eta_a*}$ & $V_2^{-\eta_b,-\eta_a}$ &
$V^{-\eta_b,\eta_a*}$ & $V^{\eta_b,-\eta_a}$ \\
\hline\hline
$D_1$ &
$\lambda_a^*\lambda_b$ & $1$ & $1$ & $1$ & $1$ \\
\hline
$A_{12}$ &
$S_4^{-\eta_a,\eta_b}$ &
$S_3^{-\eta_a,-\eta_b*}$ & $S_2^{-\eta_a,-\eta_b}$ &
$S^{-\eta_a,\eta_b*}$ & $S^{\eta_a,-\eta_b}$ \\
\hline
$B_{12}$ &
$T_4^{-\eta_a,\eta_b}$ &
$T_3^{-\eta_a,-\eta_b*}$ & $T_2^{-\eta_a,-\eta_b}$ &
$T^{-\eta_a,\eta_b*}$ & $T^{\eta_a,-\eta_b}$ \\
\hline
$C_{12}$ &
$V_4^{-\eta_a,\eta_b}$ &
$V_3^{-\eta_a,-\eta_b*}$ & $V_2^{-\eta_a,-\eta_b}$ &
$V^{-\eta_a,\eta_b*}$ & $V^{\eta_a,-\eta_b}$ \\
\hline\hline\hline
$N_2$ & 
$\eta_a$ &
$\eta_a\lambda_a^*$ & $\eta_a\lambda_b^*$ & 
$\eta_a\lambda_b$ & $\eta_a\lambda_a$ \\
\hline\hline
$A_{21}$ &
$S_4^{\eta_b,-\eta_a*}$ &
$S_2^{-\eta_b,-\eta_a}$ & $S_3^{-\eta_b,-\eta_a*}$ &
$S^{\eta_b,-\eta_a}$ & $S^{-\eta_b,\eta_a*}$ \\
\hline
$B_{21}$ &
$T_4^{\eta_b,-\eta_a*}$ &
$T_2^{-\eta_b,-\eta_a}$ & $T_3^{-\eta_b,-\eta_a*}$ &
$T^{\eta_b,-\eta_a}$ & $T^{-\eta_b,\eta_a*}$ \\
\hline
$C_{21}$ &
$V_4^{\eta_b,-\eta_a*}$ &
$V_2^{-\eta_b,-\eta_a}$ & $V_3^{-\eta_b,-\eta_a*}$ &
$V^{\eta_b,-\eta_a}$ & $V^{-\eta_b,\eta_a*}$ \\
\hline\hline
$D_2$ &
$1$ & $1$ & $1$ & $1$ & $1$ \\
\hline
$A_{22}$ &
$S_1^{-\eta_a,\eta_b*}$ &
$S_2^{-\eta_a,-\eta_b}$ & $S_3^{-\eta_a,-\eta_b*}$ &
$S^{\eta_a,-\eta_b}$ & $S^{-\eta_a,\eta_b*}$ \\
\hline
$B_{22}$ &
$T_1^{-\eta_a,\eta_b*}$ &
$T_2^{-\eta_a,-\eta_b}$ & $T_3^{-\eta_a,-\eta_b*}$ &
$T^{\eta_a,-\eta_b}$ & $T^{-\eta_a,\eta_b*}$ \\
\hline
$C_{22}$ &
$V_1^{-\eta_a,\eta_b*}$ &
$V_2^{-\eta_a,-\eta_b}$ & $V_3^{-\eta_a,-\eta_b*}$ &
$V^{\eta_a,-\eta_b}$ & $V^{-\eta_a,\eta_b*}$ \\
\hline
\end{tabular}
\label{Table:ijab2}
\end{center}
\end{table}

\vspace{10pt}

\begin{table}
\caption[]{The constant factors appearing in (\ref{eq:aibj}) for the
first five $(ai)(bj)$ Dirac neutrino scattering processes.}
\begin{center}
\begin{tabular}{|c||c|c|c|c|c|}
\hline
 & ai1 & ai2 & ai3 & ai4 & ai5 \\
\hline\hline
$N_1$ & 
$\eta_a$ & $\eta_a$ & 
$\eta_a\lambda_b^*$ & $\eta_a\lambda_a^*$ & 
$\eta_a\lambda_a$ \\
\hline\hline
$A_{11}$ &
$S_4^{\eta_b,\eta_a*}$ & $S_1^{\eta_b,\eta_a}$ &
$S_3^{-\eta_b,\eta_a*}$ & $S_2^{-\eta_b,\eta_a}$ &
$S_2^{\eta_b,-\eta_a*}$ \\
\hline
$B_{11}$ &
$T_4^{\eta_b,\eta_a*}$ & $T_1^{\eta_b,\eta_a}$ &
$T_3^{-\eta_b,\eta_a*}$ & $T_2^{-\eta_b,\eta_a}$ &
$T_2^{\eta_b,-\eta_a*}$ \\
\hline
$C_{11}$ &
$V_4^{\eta_b,\eta_a*}$ & $V_1^{\eta_b,\eta_a}$ &
$V_3^{-\eta_b,\eta_a*}$ & $V_2^{-\eta_b,\eta_a}$ &
$V_2^{\eta_b,-\eta_a*}$ \\
\hline\hline
$D_1$ &
$1$ & $\lambda_a^*\lambda_b$ & $1$ & $1$ & $1$ \\
\hline
$A_{12}$ &
$S_1^{\eta_a,\eta_b*}$ & $S_4^{\eta_a,\eta_b}$ &
$S_3^{\eta_a,-\eta_b*}$ & $S_2^{\eta_a,-\eta_b}$ &
$S_2^{-\eta_a,\eta_b*}$ \\
\hline
$B_{12}$ &
$T_1^{\eta_a,\eta_b*}$ & $T_4^{\eta_a,\eta_b}$ &
$T_3^{\eta_a,-\eta_b*}$ & $T_2^{\eta_a,-\eta_b}$ &
$T_2^{-\eta_a,\eta_b*}$ \\
\hline
$C_{12}$ &
$V_1^{\eta_a,\eta_b*}$ & $V_4^{\eta_a,\eta_b}$ &
$V_3^{\eta_a,-\eta_b*}$ & $V_2^{\eta_a,-\eta_b}$ &
$V_2^{-\eta_a,\eta_b*}$ \\
\hline\hline\hline
$N_2$ & 
$\eta_b$ & $\eta_b$ & 
$\eta_b\lambda_a^*$ & $\eta_b\lambda_b^*$ & 
$\eta_b\lambda_b$ \\
\hline\hline
$A_{21}$ &
$S_1^{\eta_b,\eta_a}$ & $S_4^{\eta_b,\eta_a*}$ &
$S_2^{-\eta_b,\eta_a}$ & $S_3^{-\eta_b,\eta_a*}$ &
$S_3^{\eta_b,-\eta_a}$ \\
\hline
$B_{21}$ &
$T_1^{\eta_b,\eta_a}$ & $T_4^{\eta_b,\eta_a*}$ &
$T_2^{-\eta_b,\eta_a}$ & $T_3^{-\eta_b,\eta_a*}$ &
$T_3^{\eta_b,-\eta_a}$ \\
\hline
$C_{21}$ &
$V_1^{\eta_b,\eta_a}$ & $V_4^{\eta_b,\eta_a*}$ &
$V_2^{-\eta_b,\eta_a}$ & $V_3^{-\eta_b,\eta_a*}$ &
$V_3^{\eta_b,-\eta_a}$ \\
\hline\hline
$D_2$ &
$\lambda_a^*\lambda_b$ & $1$ & $1$ & $1$ & $1$ \\
\hline
$A_{22}$ &
$S_4^{\eta_a,\eta_b}$ & $S_1^{\eta_a,\eta_b*}$ &
$S_2^{\eta_a,-\eta_b}$ & $S_3^{\eta_a,-\eta_b*}$ &
$S_3^{-\eta_a,\eta_b}$ \\
\hline
$B_{22}$ &
$T_4^{\eta_a,\eta_b}$ & $T_1^{\eta_a,\eta_b*}$ &
$T_2^{\eta_a,-\eta_b}$ & $T_3^{\eta_a,-\eta_b*}$ &
$T_3^{-\eta_a,\eta_b}$ \\
\hline
$C_{22}$ &
$V_4^{\eta_a,\eta_b}$ & $V_1^{\eta_a,\eta_b*}$ &
$V_2^{\eta_a,-\eta_b}$ & $V_3^{\eta_a,-\eta_b*}$ &
$V_3^{-\eta_a,\eta_b}$ \\
\hline
\end{tabular}
\label{Table:aibj1}
\end{center}
\end{table}

\begin{table}
\caption[]{The constant factors appearing in (\ref{eq:aibj}) for the
last three $(ai)(bj)$ Dirac neutrino scattering processes, and the two
Majorana neutrino ones.}
\begin{center}
\begin{tabular}{|c||c|c|c||c|c|}
\hline
 & ai6 & ai7 & ai8 & Mai1 & Mai2 \\
\hline\hline
$N_1$ & 
$\eta_a\lambda_b$ &
$\eta_a$ & $\eta_a$ & 
$\eta_a$ & $\eta_a$ \\
\hline\hline
$A_{11}$ &
$S_3^{\eta_b,-\eta_a}$ &
$S_1^{-\eta_b,-\eta_a*}$ & $S_4^{-\eta_b,-\eta_a}$ &
$S^{-\eta_b,-\eta_a*}$ & $S^{\eta_b,\eta_a}$ \\
\hline
$B_{11}$ &
$T_3^{\eta_b,-\eta_a}$ &
$T_1^{-\eta_b,-\eta_a*}$ & $T_4^{-\eta_b,-\eta_a}$ &
$T^{-\eta_b,-\eta_a*}$ & $T^{\eta_b,\eta_a}$ \\
\hline
$C_{11}$ &
$V_3^{\eta_b,-\eta_a}$ &
$V_1^{-\eta_b,-\eta_a*}$ & $V_4^{-\eta_b,-\eta_a}$ &
$V^{-\eta_b,-\eta_a*}$ & $V^{\eta_b,\eta_a}$ \\
\hline\hline
$D_1$ &
$1$ & $\lambda_a\lambda_b^*$ & $1$ & $1$ & $\lambda_a^*\lambda_b$ \\
\hline
$A_{12}$ &
$S_3^{-\eta_a,\eta_b}$ &
$S_4^{-\eta_a,-\eta_b*}$ & $S_1^{-\eta_a,-\eta_b}$ &
$S^{\eta_a,\eta_b*}$ & $S^{-\eta_a,-\eta_b}$ \\
\hline
$B_{12}$ &
$T_3^{-\eta_a,\eta_b}$ &
$T_4^{-\eta_a,-\eta_b*}$ & $T_1^{-\eta_a,-\eta_b}$ &
$T^{\eta_a,\eta_b*}$ & $T^{-\eta_a,-\eta_b}$ \\
\hline
$C_{12}$ &
$V_3^{-\eta_a,\eta_b}$ &
$V_4^{-\eta_a,-\eta_b*}$ & $V_1^{-\eta_a,-\eta_b}$ &
$V^{\eta_a,\eta_b*}$ & $V^{-\eta_a,-\eta_b}$ \\
\hline\hline\hline
$N_2$ & 
$\eta_b\lambda_a$ &
$\eta_b$ & $\eta_b$ & 
$\eta_b$ & $\eta_b$ \\
\hline\hline
$A_{21}$ &
$S_2^{\eta_b,-\eta_a*}$ &
$S_4^{-\eta_b,-\eta_a}$ & $S_1^{-\eta_b,-\eta_a*}$ &
$S^{\eta_b,\eta_a}$ & $S^{-\eta_b,-\eta_a*}$ \\
\hline
$B_{21}$ &
$T_2^{\eta_b,-\eta_a*}$ &
$T_4^{-\eta_b,-\eta_a}$ & $T_1^{-\eta_b,-\eta_a*}$ &
$T^{\eta_b,\eta_a}$ & $T^{-\eta_b,-\eta_a*}$ \\
\hline
$C_{21}$ &
$V_2^{\eta_b,-\eta_a*}$ &
$V_4^{-\eta_b,-\eta_a}$ & $V_1^{-\eta_b,-\eta_a*}$ &
$V^{\eta_b,\eta_a}$ & $V^{-\eta_b,-\eta_a*}$ \\
\hline\hline
$D_2$ &
$1$ & $1$ & $\lambda_a\lambda_b^*$ & $\lambda_a^*\lambda_b$ & $1$ \\
\hline
$A_{22}$ &
$S_2^{-\eta_a,\eta_b*}$ &
$S_1^{-\eta_a,-\eta_b}$ & $S_4^{-\eta_a,-\eta_b*}$ &
$S^{-\eta_a,-\eta_b}$ & $S^{\eta_a,\eta_b}$ \\
\hline
$B_{22}$ &
$T_2^{-\eta_a,\eta_b*}$ &
$T_1^{-\eta_a,-\eta_b}$ & $T_4^{-\eta_a,-\eta_b*}$ &
$T^{-\eta_a,-\eta_b}$ & $T^{\eta_a,\eta_b}$ \\
\hline
$C_{22}$ &
$V_2^{-\eta_a,\eta_b*}$ &
$V_1^{-\eta_a,-\eta_b}$ & $V_4^{-\eta_a,-\eta_b*}$ &
$V^{-\eta_a,-\eta_b}$ & $V^{\eta_a,\eta_b}$ \\
\hline
\end{tabular}
\label{Table:aibj2}
\end{center}
\end{table}

\vspace{10pt}

\begin{table}
\caption[]{The constant factors appearing in (\ref{eq:ajbi}) for the
first five $(aj)(bi)$ Dirac neutrino scattering processes.}
\begin{center}
\begin{tabular}{|c||c|c|c|c|c|}
\hline
 & aj1 & aj2 & aj3 & aj4 & aj5 \\
\hline\hline
$N_1$ & 
$\eta_b$ & $\eta_b$ & 
$\eta_b\lambda_a^*$ & $\eta_b\lambda_b^*$ & 
$\eta_b\lambda_b$ \\
\hline\hline
$A_{11}$ &
$S_1^{\eta_b,\eta_a}$ & $S_4^{\eta_b,\eta_a*}$ &
$S_2^{-\eta_b,\eta_a}$ & $S_3^{-\eta_b,\eta_a*}$ &
$S_3^{\eta_b,-\eta_a}$ \\
\hline
$B_{11}$ &
$T_1^{\eta_b,\eta_a}$ & $T_4^{\eta_b,\eta_a*}$ &
$T_2^{-\eta_b,\eta_a}$ & $T_3^{-\eta_b,\eta_a*}$ &
$T_3^{\eta_b,-\eta_a}$ \\
\hline
$C_{11}$ &
$V_1^{\eta_b,\eta_a}$ & $V_4^{\eta_b,\eta_a*}$ &
$V_2^{-\eta_b,\eta_a}$ & $V_3^{-\eta_b,\eta_a*}$ &
$V_3^{\eta_b,-\eta_a}$ \\
\hline\hline
$D_1$ &
$\lambda_a^*\lambda_b$ & $1$ & $1$ & $1$ & $1$ \\
\hline
$A_{12}$ &
$S_4^{\eta_a,\eta_b}$ & $S_1^{\eta_a,\eta_b*}$ &
$S_2^{\eta_a,-\eta_b}$ & $S_3^{\eta_a,-\eta_b*}$ &
$S_3^{-\eta_a,\eta_b}$ \\
\hline
$B_{12}$ &
$T_4^{\eta_a,\eta_b}$ & $T_1^{\eta_a,\eta_b*}$ &
$T_2^{\eta_a,-\eta_b}$ & $T_3^{\eta_a,-\eta_b*}$ &
$T_3^{-\eta_a,\eta_b}$ \\
\hline
$C_{12}$ &
$V_4^{\eta_a,\eta_b}$ & $V_1^{\eta_a,\eta_b*}$ &
$V_2^{\eta_a,-\eta_b}$ & $V_3^{\eta_a,-\eta_b*}$ &
$V_3^{-\eta_a,\eta_b}$ \\
\hline\hline\hline
$N_2$ & 
$\eta_a$ & $\eta_a$ & 
$\eta_a\lambda_b^*$ & $\eta_a\lambda_a^*$ & 
$\eta_a\lambda_a$ \\
\hline\hline
$A_{21}$ &
$S_4^{\eta_b,\eta_a*}$ & $S_1^{\eta_b,\eta_a}$ &
$S_3^{-\eta_b,\eta_a*}$ & $S_2^{-\eta_b,\eta_a}$ &
$S_2^{\eta_b,-\eta_a*}$ \\
\hline
$B_{21}$ &
$T_4^{\eta_b,\eta_a*}$ & $T_1^{\eta_b,\eta_a}$ &
$T_3^{-\eta_b,\eta_a*}$ & $T_2^{-\eta_b,\eta_a}$ &
$T_2^{\eta_b,-\eta_a*}$ \\
\hline
$C_{21}$ &
$V_4^{\eta_b,\eta_a*}$ & $V_1^{\eta_b,\eta_a}$ &
$V_3^{-\eta_b,\eta_a*}$ & $V_2^{-\eta_b,\eta_a}$ &
$V_2^{\eta_b,-\eta_a*}$ \\
\hline\hline
$D_2$ &
$1$ & $\lambda_a^*\lambda_b$ & $1$ & $1$ & $1$ \\
\hline
$A_{22}$ &
$S_1^{\eta_a,\eta_b*}$ & $S_4^{\eta_a,\eta_b}$ &
$S_3^{\eta_a,-\eta_b*}$ & $S_2^{\eta_a,-\eta_b}$ &
$S_2^{-\eta_a,\eta_b*}$ \\
\hline
$B_{22}$ &
$T_1^{\eta_a,\eta_b*}$ & $T_4^{\eta_a,\eta_b}$ &
$T_3^{\eta_a,-\eta_b*}$ & $T_2^{\eta_a,-\eta_b}$ &
$T_2^{-\eta_a,\eta_b*}$ \\
\hline
$C_{22}$ &
$V_1^{\eta_a,\eta_b*}$ & $V_4^{\eta_a,\eta_b}$ &
$V_3^{\eta_a,-\eta_b*}$ & $V_2^{\eta_a,-\eta_b}$ &
$V_2^{-\eta_a,\eta_b*}$ \\
\hline
\end{tabular}
\label{Table:ajbi1}
\end{center}
\end{table}

\begin{table}
\caption[]{The constant factors appearing in (\ref{eq:ajbi}) for the
last three $(aj)(bi)$ Dirac neutrino scattering processes, and the
two Majorana neutrino ones.}
\begin{center}
\begin{tabular}{|c||c|c|c||c|c|}
\hline
 & aj6 & aj7 & aj8 & Maj1 & Maj2 \\
\hline\hline
$N_1$ & 
$\eta_b\lambda_a$ &
$\eta_b$ & $\eta_b$ & 
$\eta_b$ & $\eta_b$ \\
\hline\hline
$A_{11}$ &
$S_2^{\eta_b,-\eta_a*}$ &
$S_4^{-\eta_b,-\eta_a}$ & $S_1^{-\eta_b,-\eta_a*}$ &
$S^{\eta_b,\eta_a}$ & $S^{-\eta_b,-\eta_a*}$ \\
\hline
$B_{11}$ &
$T_2^{\eta_b,-\eta_a*}$ &
$T_4^{-\eta_b,-\eta_a}$ & $T_1^{-\eta_b,-\eta_a*}$ &
$T^{\eta_b,\eta_a}$ & $T^{-\eta_b,-\eta_a*}$ \\
\hline
$C_{11}$ &
$V_2^{\eta_b,-\eta_a*}$ &
$V_4^{-\eta_b,-\eta_a}$ & $V_1^{-\eta_b,-\eta_a*}$ &
$V^{\eta_b,\eta_a}$ & $V^{-\eta_b,-\eta_a*}$ \\
\hline\hline
$D_1$ &
$1$ & $1$ & $\lambda_a\lambda_b^*$ & $\lambda_a^*\lambda_b$ & $1$ \\
\hline
$A_{12}$ &
$S_2^{-\eta_a,\eta_b*}$ &
$S_1^{-\eta_a,-\eta_b}$ & $S_4^{-\eta_a,-\eta_b*}$ &
$S^{-\eta_a,-\eta_b}$ & $S^{\eta_a,\eta_b*}$ \\
\hline
$B_{12}$ &
$T_2^{-\eta_a,\eta_b*}$ &
$T_1^{-\eta_a,-\eta_b}$ & $T_4^{-\eta_a,-\eta_b*}$ &
$T^{-\eta_a,-\eta_b}$ & $T^{\eta_a,\eta_b*}$ \\
\hline
$C_{12}$ &
$V_2^{-\eta_a,\eta_b*}$ &
$V_1^{-\eta_a,-\eta_b}$ & $V_4^{-\eta_a,-\eta_b*}$ &
$V^{-\eta_a,-\eta_b}$ & $V^{\eta_a,\eta_b*}$ \\
\hline\hline\hline
$N_2$ & 
$\eta_a\lambda_b$ &
$\eta_a$ & $\eta_a$ & 
$\eta_a$ & $\eta_a$ \\
\hline\hline
$A_{21}$ &
$S_3^{\eta_b,-\eta_a}$ &
$S_1^{-\eta_b,-\eta_a*}$ & $S_4^{-\eta_b,-\eta_a}$ &
$S^{-\eta_b,-\eta_a*}$ & $S^{\eta_b,\eta_a}$ \\
\hline
$B_{21}$ &
$T_3^{\eta_b,-\eta_a}$ &
$T_1^{-\eta_b,-\eta_a*}$ & $T_4^{-\eta_b,-\eta_a}$ &
$T^{-\eta_b,-\eta_a*}$ & $T^{\eta_b,\eta_a}$ \\
\hline
$C_{21}$ &
$V_3^{\eta_b,-\eta_a}$ &
$V_1^{-\eta_b,-\eta_a*}$ & $V_4^{-\eta_b,-\eta_a}$ &
$V^{-\eta_b,-\eta_a*}$ & $V^{\eta_b,\eta_a}$ \\
\hline\hline
$D_2$ &
$1$ & $\lambda_a\lambda_b^*$ & $1$ & $1$ & $\lambda_a^*\lambda_b$ \\
\hline
$A_{22}$ &
$S_3^{-\eta_a,\eta_b}$ &
$S_4^{-\eta_a,-\eta_b*}$ & $S_1^{-\eta_a,-\eta_b}$ &
$S^{\eta_a,\eta_b*}$ & $S^{-\eta_a,-\eta_b}$ \\
\hline
$B_{22}$ &
$T_3^{-\eta_a,\eta_b}$ &
$T_4^{-\eta_a,-\eta_b*}$ & $T_1^{-\eta_a,-\eta_b}$ &
$T^{\eta_a,\eta_b*}$ & $T^{-\eta_a,-\eta_b}$ \\
\hline
$C_{22}$ &
$V_3^{-\eta_a,\eta_b}$ &
$V_4^{-\eta_a,-\eta_b*}$ & $V_1^{-\eta_a,-\eta_b}$ &
$V^{\eta_a,\eta_b*}$ & $V^{-\eta_a,-\eta_b}$ \\
\hline
\end{tabular}
\label{Table:ajbi2}
\end{center}
\end{table}

\end{document}